\documentclass[prb,twocolumn,superscriptaddress,citeautoscript]{revtex4}
\usepackage{amssymb}
\usepackage{amsmath}
\usepackage{amsfonts}
\usepackage{color}
\usepackage{graphicx}
\usepackage{bm}
\newcommand* {\vek}[1]{{\ensuremath{\bm{\mathrm{#1}}}}}
\newcommand* {\vekc}[1]{{\ensuremath{\bm{\mathcal{#1}}}}}
\newcommand* {\kk}{\vek{k}}

\newcommand* {\Ds}{\displaystyle}

\newcommand* {\braket}[1]{\ensuremath{\langle {#1} \rangle}}
\usepackage{array}
\newcolumntype {s}[1]{@{\hspace{#1}}} 
\newcolumntype {R}{>{$}r<{$}}         
\newcolumntype {C}{>{$}c<{$}}         
\newcolumntype {L}{>{$}l<{$}}         
\usepackage{dcolumn}

\newcommand* {\kcomp}{\kappa}
\newcommand* {\kvek}{\bm{\kcomp}}

\newcommand* {\strain}{\epsilon}
\newcommand* {\Strain}{\vek{\epsilon}}
\newcommand* {\Ee}{\mathcal{E}}
\newcommand* {\koeff}[3]{#1^{#2}_{#3}}

\newenvironment{textmath}{$\displaystyle}{$}

\newcommand{\allowed}[1]{\mbox{$\bm{#1}$}}
\newcommand{\allowednew}[1]{\allowed{\textcolor{red}{#1}}}
\newcommand{\forbidden}[1]{\mbox{$#1$}}
\newcommand{\newinblg}[1]{\relax\ifmmode#1\else#1$^\star$\fi}

\graphicspath{{.}{./EPS/}}

\begin{document}
\title{Magneto-electric equivalence and emergent electrodynamics
in bilayer graphene}

\author{R. Winkler}
\affiliation{Department of Physics, Northern Illinois University,
DeKalb, Illinois 60115, USA}
\affiliation{Materials Science Division, Argonne National
Laboratory, Argonne, Illinois 60439, USA}
\affiliation{Department of Physical Chemistry,
The University of the Basque Country, 48080 Bilbao, Spain}
\affiliation{IKERBASQUE, Basque Foundation for Science, 48011
  Bilbao, Spain}

\author{U. Z\"ulicke}
\affiliation{School of Chemical and Physical Sciences and
MacDiarmid Institute for Advanced Materials and Nanotechnology,
Victoria University of Wellington, PO Box 600, Wellington 6140,
New Zealand}

\date{Version of \today. Not for distribution!}

\begin{abstract}
It is a fundamental paradigm that the physical effects induced by
electric fields are qualitatively different from those induced by
magnetic fields. Here we show that electrons at a Dirac point in bilayer
graphene experience an unusual type of electromagnetism where
magnetic and electric fields are virtually equivalent: every coupling
of an electron's degrees of freedom to a \emph{magnetic\/} field is
matched by an analogous coupling of the \emph{same} degrees of
freedom to an \emph{electric\/} field. This counter-intuitive duality of
matter-field interactions enables novel ways to create and manipulate
spin and pseudo-spin polarizations in bilayer graphene via external
fields and leads to the emergence of a valley-contrasting axion
electrodynamics, where the traditional association of charges at rest
with electric fields and charge currents with magnetic fields is reversed.
\end{abstract}

\maketitle

\section{Introduction}

\begin{figure}[b]
  \includegraphics[width=0.9\columnwidth]{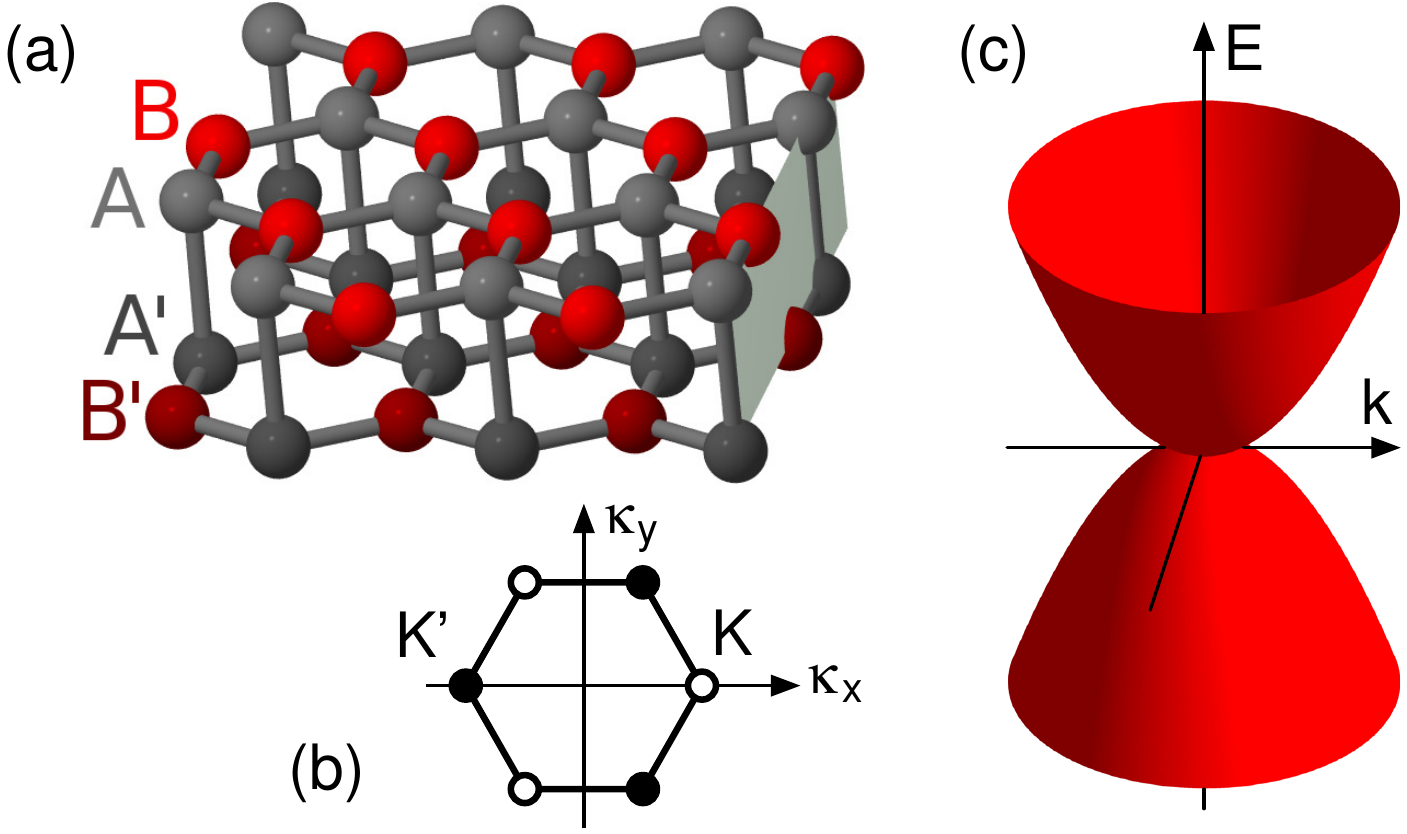}
  \caption{\label{fig:lattice}Basic structural and electronic properties
  of bilayer graphene. (a) Honeycomb structure of a bilayer
  stack of graphene. Atoms in sublattice $A$ ($B$) are marked in
  grey (red). A $yz$ plane is marked in light grey. (b) Brillouin
  zone and its two inequivalent corner points $\vek{K}$ and
  $\vek{K}'$. The remaining corners are related with $\vek{K}$ or
  $\vek{K}'$ by reciprocal lattice vectors. (c) Dispersion $E(k)$
  near the $\vek{K}$ point. We have $\kk \equiv \kvek - \vek{K}$.}
\end{figure}

The dynamics of charge carriers in crystalline solids normally
resembles that exhibited by free electrons, only with numerical
values of parameters such as mass and gyromagnetic ratio
renormalized due to the influence of the crystal
structure~\cite{ros09}. In certain instances, changes in the
properties of band electrons turn out to be dramatic. Few-layer
samples of graphite~\cite{gei07, cas09} are one example, and a new
class of so-called \emph{topological\/} materials~\cite{moo09}
another. Here we focus on bilayer graphene~\cite{nov06, mcc06}
(BLG), i.e., two single-layer sheets of graphite stacked as shown in
Fig.~\ref{fig:lattice}(a). Electrons in this material turn out to be
pseudospin-carrying chiral fermions having a finite band mass but
zero rest energy~\cite{gei09}. Our study shows that the interaction
of these exotic charge carriers with electromagnetic fields is very
unusual. Normally, electric fields couple to electric charges at
rest and moving magnetic moments, whereas magnetic fields couple to
moving electric charges and magnetic moments at rest. In contrast,
every coupling of a bilayer electron's degree of freedom to an
electric field is matched by an analogous coupling to a magnetic
field. We explore physical consequences of this magneto-electric
equivalence, including anomalous polarizations of real spin and the
layer-index-related pseudospin, as well as an anisotropic version of
axion electrodynamics~\cite{wil87}.

The BLG band structure near the $\vek{K}$ point in the Brillouin
zone [see Fig.~\ref{fig:lattice}(b)] is described by the effective
Hamiltonian
\begin{eqnarray}\label{eq:effM_BLG}
\mathcal{H}_\vek{K} (\kk) &=& -\frac{\hbar^2}{2 m_1} \left(k_+^2 \,
\sigma_+ + k_-^2 \, \sigma_-\right) + \frac{\hbar^2 k^2}{2 m_2}\,
\sigma_0 \nonumber \\ && \hspace{2cm} - \hbar v_3 \left( k_- \, \sigma_+
+ k_+ \, \sigma_-\right ) \, ,
\end{eqnarray}
where $\hbar$ is Planck's constant, $\vek{k}\equiv(k_x, k_y)$ is the
electrons' wave vector measured from $\vek{K}$, and the Pauli
matrices $\sigma_{x,y,z}$ are associated with the sublattice (or,
equivalently, the layer-index) pseudospin degree of
freedom~\cite{cas09}. In our notation, $\sigma_0$ is the $2\times 2$
unit matrix, $\sigma_\pm=(\sigma_x \pm i\sigma_y)/2$, and
$k_\pm=k_x\pm i k_y$. Numerical values for the effective masses
$m_j$ and the speed $v_3$ are well known~\cite{mcc06, cas09}. Very
close to the $\vek{K}$ point, the energy dispersion resulting from
(\ref{eq:effM_BLG}) mimics that of massless Dirac electrons, as is
the case in single-layer graphene (SLG). However, as $m_1\ll m_2$,
the dominant behavior of electrons in BLG is captured by the quadratic
dispersion shown in Fig.~\ref{fig:lattice}(c).

\begin{table*}[t] \caption{\label{tab:newTerms} Magneto-electric
  equivalence of matter-field interactions in BLG. In each row,
  the coupling of an electron's degrees of freedom to a magnetic field
  $\vek{B}$ (left) is matched by an analogous coupling to an
  electric field $\vekc{E}$ (right). In these pairs, $\vek{B}$ and
  $\vekc{E}$ couple identically to the real spin $\vek{s}$ and pseudospin
  $\vek{\sigma}$, while the coupling to the valley isospin $\vek{\tau}$ is
  opposite: For one field, the coupling has the same sign in both valleys
  (isospin $\tau_0$), for the other field it has opposite signs
  (isospin $\tau_z$). In each row, the interaction unique to BLG is marked
  by $\star\,$, while the other one exists also in SLG~\cite{win10a}.
  The interactions are visualized in Fig.~\ref{fig:visual}.}
  \renewcommand{\arraystretch}{1.2}
  \begin{tabular}{rRs{1em}cs{1em}Ll} \hline \hline
\multicolumn{2}{c}{magnetic field $\vek{B}$} &
& \multicolumn{2}{c}{electric field $\vekc{E}$} \\ \hline
orbital Zeeman splitting ($\perp$ field) &
B_z\,\sigma_z\,\tau_z &
(1) &
\newinblg{\Ee_z\,\sigma_z\,\tau_0} &
\newinblg{inter-layer (pseudo-spin) gap} \\
magnetic spin splitting ($\perp$ field) &
B_z\, s_z\,\tau_0 &
(2) &
\newinblg{\Ee_z\, s_z\, \tau_z} &
\newinblg{electric spin splitting ($\perp$ field)} \\
magnetic spin splitting ($\parallel$ field) &
(B_x\, s_x + B_y\, s_y)\tau_0  &
(3) &
\newinblg{(\Ee_x\, s_x + \Ee_y\, s_y)\tau_z} &
\newinblg{electric spin splitting ($\parallel$ field)} \\
\newinblg{spin-orbital Zeeman ($\perp$ field)} &
\newinblg{B_z (s_y \, \sigma_x\,\tau_0 - s_x \, \sigma_y\,\tau_z)} &
(4) &
\Ee_z (s_y \, \sigma_x\,\tau_z - s_x \, \sigma_y\,\tau_0) &
Rashba spin splitting ($\perp$ field) \\
\newinblg{spin-orbital Zeeman ($\parallel$ field)} &
\newinblg{s_z ( B_y\, \sigma_x\,\tau_0 - B_x\, \sigma_y\,\tau_z)} &
(5) &
s_z (\Ee_y \, \sigma_x\,\tau_z - \Ee_x \, \sigma_y\,\tau_0 ) &
Rashba spin splitting ($\parallel$ field) \\
\newinblg{orbital Zeeman splitting ($\parallel$ field)} &
\newinblg{(k_x\, B_y - k_y\, B_x)  \sigma_z \,\tau_0} &
(6) &
(k_x\,\Ee_y - k_y\, \Ee_x) \sigma_z\,\tau_z &
orbital Rashba splitting \\
trigonal Zeeman splitting &
\begin{array}[t]{@{}r@{}}
  (s_y\, B_y - s_x\, B_x) \sigma_x\,\tau_0 \\  
  + (s_x\, B_y + s_y \, B_x) \sigma_y\, \tau_z
\end{array} &
(7) &
\newinblg{\begin{array}[t]{@{}l@{}}
  (s_y\,\Ee_y - s_x\, \Ee_x) \sigma_x\,\tau_z \\
  +(s_x\, \Ee_y + s_y\, \Ee_x)\sigma_y\,\tau_0
\end{array}} &
\newinblg{trigonal Rashba splitting} \\
\hline \hline
\end{tabular}
\end{table*}

External fields can have striking effects on the electronic
properties of charge carriers in solids. Previously, only the
effects of electric fields $\Ee_z$ and magnetic fields $B_z$
directed perpendicular to the BLG sheet have been considered, giving
the extended effective Hamiltonian
\begin{equation}\label{eq:fieldHam}
  \begin{array}[b]{l}
    \mathcal{H}_\vek{K} (\kk, \Ee_z, B_z)
    = \mathcal{H}_\vek{K} (\kk+e\vek{A}) \\[0.5ex] \hspace{3em}
    \Ds {} + \frac{\lambda_z}{2}\, \Ee_z\, \sigma_z 
    - \frac{g}{2}\,\mu_\mathrm{B}\,B_z\,\sigma_z
    + c_z \,\Ee_z B_z\, \sigma_0 \, . \hspace{1em}
  \end{array}
\end{equation}
Accordingly, (i)~a potential difference between the two layers
(equivalent to finite $\Ee_z$) opens up a pseudospin
gap~\cite{mcc06, oht06} $\lambda_z \Ee_z$, (ii)~$B_z$ induces a
pseudospin Zeeman splitting~\cite{zha11} $g \,\mu_\mathrm{B}\,B_z$,
and (iii)~the simultaneous presence of fields $\Ee_z$ and $B_z$
leads to an (actually, valley-contrasting -- see below) overall
energy shift~\cite{nak09, kos10, zha11} $c_z \,\Ee_z B_z$. In
Eq.~(\ref{eq:fieldHam}), $g$ is the gyromagnetic ratio,
$\mu_\mathrm{B}$ is the Bohr magneton, and $\vek{A}$ is the
electromagnetic vector potential satisfying $(\vek{\nabla} \times
\vek{A})_z = B_z$. The matter-field interactions (i)--(iii) generate
sizeable effects for typical values of $\Ee_z$ and $B_z$ (see
Appendix~\ref{app:C}). In addition, the chiral nature of electrons
in BLG leads to unconventional quantum-Hall physics~\cite{nov06, mcc06}.

\section{Magneto-electric equivalence in BLG: Origin \& physical
consequences}

Inspection of Eq.~(\ref{eq:fieldHam}) reveals a surprising feature:
disregarding constant prefactors, the electron's interaction with
fields $\Ee_z$ and $B_z$ is symmetric with respect to the
interchange of $\Ee_z$ and $B_z$. Indeed, this observation is not an
accident. It reflects the unusual property of BLG that its crystal
symmetry does not distinguish between polar vectors such as the
electric field $\vekc{E}$ and axial vectors such as the magnetic
field $\vek{B}$. Moreover, the familiar constraints due to
time-reversal invariance are modified in BLG such that symmetry
under time reversal likewise permits that $\vekc{E}$ and $\vek{B}$
become interchangeable.

Using symmetry\cite{bir74, win03, win10a} we have obtained the
complete set of interactions of electrons near the $\vek{K}$ point
in BLG with magnetic and electric fields $\vek{B}$ and $\vekc{E}$,
strain $\Strain$, and also spin $\vek{s}$. See Appendix~\ref{app:A}
for details on our method and full results. All
interactions that exist in SLG are also present in BLG. In addition,
there are new interactions in BLG that can be obtained from
field-dependent SLG interactions by replacing the magnetic
(electric) field components with their electric (magnetic)
counterparts. Table~\ref{tab:newTerms} summarizes this
magneto-electric equivalence for couplings that are linear in
electric or magnetic field components. The interaction $\propto
\Ee_z \sigma_z$ coupling the pseudospin to a perpendicular electric
field thus arises by magneto-electric equivalence from the
pseudospin Zeeman term $\propto B_z\sigma_z$ that exists in SLG.
Moreover, we obtain a rather counter-intuitive purely
electric-field-dependent spin splitting $\propto \Ee_z s_z$, which
can be directly measured using electron spin resonance. Also
interesting is the spin-orbital Zeeman splitting $\propto B_z (s_y
\, \sigma_x - s_x \, \sigma_y)$, which is the magneto-electric
equivalent of Rashba spin splitting in SLG~\cite{kan05a}. We note
that the electrons interact not only with the fields $\vek{B}$ and
$\vekc{E}$, but their quantum dynamics depends also -- via the usual
minimal coupling~\cite{ros09} -- on the electromagnetic potentials
$\vek{A}$ and $\Phi$. The latter contributions are not affected by
the magneto-electric equivalence. 

\subsection{Spin textures}

\begin{figure*}
  \begin{tabular}{cc}
    & \makebox[0.30\textwidth]{valley $\vek{K}: \vek{B}, \vekc{E}$}%
      \makebox[0.30\textwidth]{valley $\vek{K}': \vek{B}$}%
      \makebox[0.30\textwidth]{valley $\vek{K}': \vekc{E}$} \\
      \raisebox{13ex}[0pt]{\renewcommand{\arraystretch}{8.15}
       \begin{tabular}[b]{c}
         (0) \\ (1) \\ (4) \\ (5) \\ (6) \\ (7)
       \end{tabular}}
    & \includegraphics[width=0.90\textwidth]{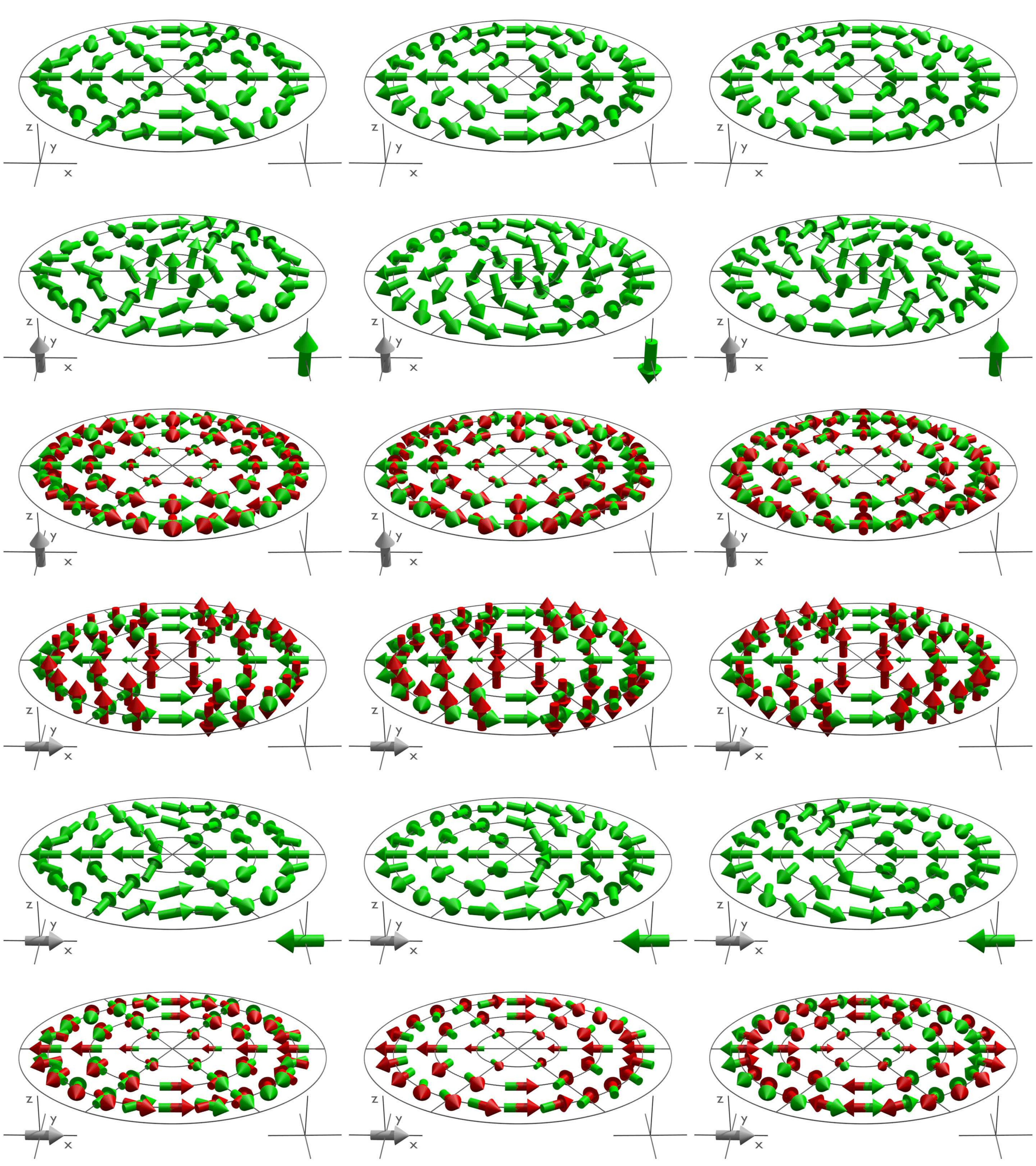}
  \end{tabular}
  \caption{\label{fig:visual}Visualization of real and pseudo-spin
  textures in BLG generated by field-dependent interactions shown in
  Table~\ref{tab:newTerms} [omitting the trivial interactions (2) and (3)].
  For selected points in $\vek{k}$ space, a red (green) arrow
  indicates the expectation value of the real (pseudo-)spin vector
  of the corresponding negative-energy eigenstate obtained
  by diagonalizing the leading field-independent contribution to the
  BLG Hamiltonian together with the interaction associated with the same
  number in Table~\ref{tab:newTerms}. The 1st (2nd, 3rd) column
  shows results for the $\vek{K}$ ($\vek{K'}$ with $\vek{B}$, $\vek{K'}$
  with $\vekc{E}$) valley. Note that our sign convention is such that
  $\vek{B}$ and $\vekc{E}$ have the same effect in the $\vek{K}$
  valley (1st column). Grey arrows indicate the direction of the
  applied field. The first row denoted (0) shows the pseudo-spin
  orientation without any external fields. Big arrows in the lower
  right of the panels indicate the existence of a net spin
  polarization obtained by averaging over the (pseudo-)spin
  expectation values of occupied hole states for a negative chemical
  potential.}
\end{figure*}

The interactions shown in
Table~\ref{tab:newTerms} give rise to a diverse set of textures for
the induced spin and pseudospin orientations in $\kk$ space, which
are illustrated in Fig.~\ref{fig:visual}. The grey arrow in the
lower left of each panel indicates the direction of the external
field while a large arrow in the lower right indicates (if present)
the spin polarization induced by the particular interaction when
averaging over all occupied states in the $k_xk_y$ plane.

In addition to the layer-index pseudospin $\vek{\sigma}$, electrons
in BLG carry a valley-isospin $\vek{\tau}$ that distinguishes states
near the two inequivalent $\vek{K}$ and $\vek{K}'\equiv -\vek{K}$
points in the Brillouin zone~\cite{cas09} [Fig.~\ref{fig:lattice}(b)]. The
effective Hamiltonian for electrons in the $\vek{K}'$ valley can be
obtained from that for the $\vek{K}$ valley by a reflection of the
vectors $\kk$, $\vek{s}$, $\vekc{E}$, and $\vek{B}$ at the $yz$
plane, see Fig.~\ref{fig:lattice}(a)~\cite{win10a}. Choosing the
convention that the $\vek{B}$ and $\vekc{E}$-dependent terms have
the same sign in the $\vek{K}$ valley (first column of
Fig.~\ref{fig:visual}), the corresponding term in the $\vek{K}'$
valley involving the axial vector $\vek{B}$ (second column) differs
by an overall minus sign from the term involving the polar vector
$\vekc{E}$ (third column of Fig.~\ref{fig:visual}, see also
Table~\ref{tab:newTerms}). For example, the orbital Zeeman term
$\propto B_z \sigma_z\tau_z$ couples $B_z$ to the pseudospin
associated with the sublattice degree of freedom $\sigma_z$.
However, the field $B_z$ does not induce a global pseudo-spin
(sublattice) polarization, because this term has opposite signs in
the two valleys so that the pseudospin polarization in the two
valleys is antiparallel. As a result, the two sublattices remain
indistinguishable even for finite $B_z$. Only an electric field
$\Ee_z$ polarizes the sublattices in BLG via the term
$\propto\Ee_z\sigma_z\tau_0$, consistent with the fact that the
pseudospin $\sigma_z$ is even under time reversal~\cite{win10}. This
behaviour is opposite to that of real spin, where $B_z s_z \tau_0$
induces a real-spin polarization $\braket{s_z}$ (when averaging over
the occupied states in both valleys) while the term $\Ee_z s_z
\tau_z$ does not, consistent with time reversal symmetry. Similarly,
an in-plane magnetic field $\vek{B}_\|$ gives rise to a macroscopic
in-plane polarization of real spins, whereas the real-spin
polarization induced by an electric field $\vekc{E}_\|$ is
anti-parallel in the two valleys.

The fields $\vek{B}_\|$ and $\vekc{E}_\|$ also couple to the
in-plane pseudospin $\vek{\sigma}_\|$. Term (6) of
Table~\ref{tab:newTerms} induces an out-of-plane tilt of the spin
orientation of individual states which, in the $\vek{K}'$ valley, has
opposite signs for $\vek{B}_\|$ and $\vekc{E}_\|$. Remarkably, on
average this yields an in-plane polarization
$\braket{\vek{\sigma}_\|}$ which is nonetheless the same in each
valley for fields $\vek{B}_\|$ and $\vekc{E}_\|$
(Fig.~\ref{fig:visual}). This result reflects the fact that the
macroscopic pseudospin polarization $\braket{\vek{\sigma}_\|}$ is
neither even nor odd under time reversal~\cite{win10}. More
precisely, in each valley the direction of
$\braket{\vek{\sigma}_\|}$ is well-defined only up to a
gauge-dependent angular offset. Yet the \emph{change} of
$\braket{\vek{\sigma}_\|}$ induced by a change in the in-plane
orientation of the applied field \emph{is} well-defined and it
points clockwise in one valley and counterclockwise in the other
valley (for both $\vek{B}_\|$ and $\vekc{E}_\|$ and all terms in
Tab.~\ref{tab:newTerms} giving rise to an in-plane pseudospin
orientation of individual states). Specifically for the term (6), if
we change the in-plane orientation of the fields $\vek{B}_\|$ or
$\vekc{E}_\|$ by an angle $\varphi$, this changes the resulting
average polarization $\braket{\vek{\sigma}_\|}$ by $\pm 2\varphi$.
This implies, in particular, that reverting the direction of the
external field yields the same orientation of the induced pseudospin
polarization. We see here that the pseudospin polarization induced
by external fields $\vekc{E}$ and $\vek{B}$ behaves qualitatively
different from the polarization of real spins.

Numeric values for the prefactors of the field-de\-pen\-dent
interactions in Fig.~\ref{fig:visual} cannot be determined with the
methods used here, but the important qualitative trends do not
depend on these values. Therefore, in the calculations yielding
Fig.~\ref{fig:visual}, the prefactors were chosen large enough to
provide a visually clear picture.

\subsection{Valley-contrasting axion electrodynamics}

Up to now, we have
considered implications of the magneto-electric equivalence for
interactions that are linear in either $\vek{B}$ or $\vekc{E}$. In
every material, we also have interactions proportional to the
squared field components. In SLG, these are interactions
proportional to $B_\|^2$, $B_z^2$, $\Ee_\|^2$ and $\Ee_z^2$. The
magneto-electric equivalence implies that we may replace one factor
in the square of one field by the corresponding components of the
other field.
We thus get scalar magneto-electric couplings
\begin{equation}
    \label{eq:me_scalar}
    c_\| \vekc{E}_\| \cdot \vek{B}_\| \tau_z
    + c_z\, \Ee_z\, B_z \,\tau_z \, ,
\end{equation}
as well as their trigonally anisotropic counterpart
\begin{equation}
    \label{eq:me_gauge}
    c_\bigtriangleup [(\Ee_y \, B_y - \Ee_x\, B_x)\sigma_x\,\tau_z
    + (\Ee_y\, B_x + \Ee_x\, B_y)\sigma_y\,\tau_0] \, .
\end{equation}
Such scalars enter not only the Hamiltonian describing the dynamics
of electrons in graphene, but they also enter the
Lagrangian~\cite{jac99} from which we obtain Maxwell's equations in
graphene.
The bi-linear coupling of electric and magnetic fields displayed in
Eq.~(\ref{eq:me_scalar}) is reminiscent of the $\vek{\Ee}\cdot
\vek{B}$ contribution to the Lagrangian of axions~\cite{wil87},
hypothetical particles that were introduced to solve the strong CP
problem in particle physics. Here the bi-linear couplings
also have a valley dependence. Assuming independent dynamics for the
electrons in the two valleys, the new term in the Lagrangian becomes
\begin{equation}
  \mathcal{L}_\mathrm{ax} = \sum_{\alpha=\vek{K}, \vek{K}'}
  (- \nu_\alpha) \left( c_\| \vek{\Ee}_\|\cdot\vek{B}_\|
    + c_z\, \Ee_z\, B_z \right) ,
\end{equation}
where $\nu_{\vek{K}} = +1$ and $\nu_{\vek{K}'} = -1$. The presence
of $\mathcal{L}_\mathrm{ax}$ modifies Gauss' law in BLG by inducing
valley-dependent charge densities
\begin{equation}\label{eq:axion_rho}
  \rho^{(\alpha)}  =
  \nu_\alpha \left[ \vek{B}_\| \cdot \vek{\nabla}_\| c_\|
    + \left(c_\| -  c_z \right) \vek{\nabla}_\| \cdot \vek{B}_\| \right]
\end{equation}
proportional to $\vek{B}_\|$, and it modifies Amp\`ere's law by
inducing current densities
\begin{equation}\label{eq:axion_j}
  \vek{j}^{(\alpha)}_\| = \nu_\alpha \,\, \hat{\vek{z}}\times
  \left[  \Ee_z \vek{\nabla}_\| c_z
    + \left(c_z - c_\| \right) \vek{\nabla}_\| \Ee_z \right]
\end{equation}
proportional to $\Ee_z$ (see Appendix~\ref{app:B} for
details), thus realizing a valley-contrasting axion electrodynamics
in BLG, where the usual coupling~\cite{jac99} of charges to electric
fields (Gauss' law) and currents to magnetic fields (Amp\`ere's law)
is completely reversed.

\begin{figure}[t]
  \includegraphics[width=1.0\columnwidth]{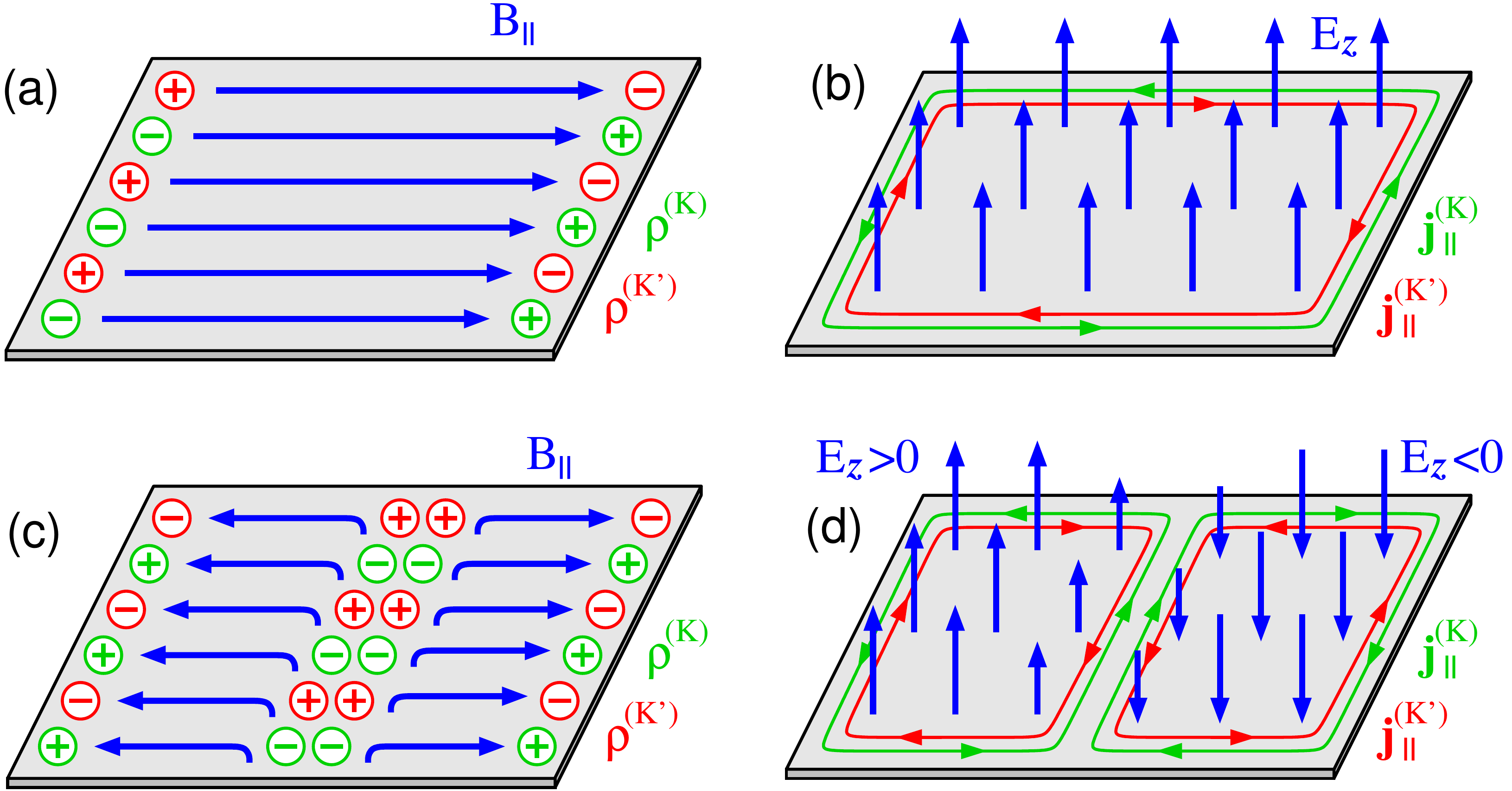}
  \caption{\label{fig:axion}Ramifications of the valley-contrasting
  axion electrodynamics. (a) An in-plane magnetic field $\vek{B}_\|$
  induces equilibrium valley-isospin densities $\rho^{(\alpha)}$ at
  sample boundaries that are not aligned with $\vek{B}_\|$.
  (b)~A perpendicular electric field $\Ee_z$ gives rise to
  equilibrium valley-isospin edge currents $\vek{j}_\|^{(\alpha)}$.
  (c)~Valley-isospin densities can be created inside the sample by
  engineering inhomogeneity in the in-plane magnetic field.
  (d)~Valley-isospin currents can be created inside the sample by
  engineering inhomogeneity in the perpendicular electric field.}
\end{figure}

The finiteness of any real sample leads to a spatial variation of
the coefficients $c_{j}$ at its edge. As a result, equilibrium
valley-isospin \emph{densities\/} are induced at the BLG sheet's
boundaries in the $xy$ plane when the magnetic field has an in-plane
component directed perpendicularly to such a boundary; and a
perpendicular electric field generates equilibrium valley-isospin
\emph{currents\/} flowing parallel to the system's boundaries in the
$xy$ plane. Figure~\ref{fig:axion} illustrates these effects in
panels (a) and (b).
The valley-asymmetric character of the edge densities and currents
in BLG distinguishes them from similar edge effects in topological
materials~\cite{qi08} that generally require the presence of strong
spin-orbit coupling. Also in contrast to the usually
considered~\cite{wil87, qi08} cases where $c_z = c_\|$,
valley-isospin densities (currents) can be induced in the
\emph{bulk\/} of BLG by spatially inhomogeneous fields $\vek{B}_\|$
($\Ee_z$). Scenarios for this are illustrated in panels (c) and (d)
of Fig.~\ref{fig:axion}, which can be realized, e.g., by placing
nanomagnets on the surface of a BLG sample [for (c)] or arranging
pairs of front and back gates
with opposite voltage polarity around adjacent parts of
the sample~\cite{mar08} [for (d)]. More generally, the
valley-contrasting axion electrodynamics discovered here enlarges
the scope of valley-dependent electronic effects in graphene
materials~\cite{xia07, ryc07} by providing a comprehensive framework
to design magneto-electric effects at boundaries and interfaces in
BLG, including associated valley-helical edge states~\cite{mar08,
li11a}.

\section{Discussion and outlook}

One of the most exciting aspects of magneto-electric equivalence in
BLG is the wide range of novel matter-field interactions resulting
in this material. To achieve any type of effect within a
valley, either an electric or a magnetic field component can be
utilized. This includes, in particular, the different real-spin
and pseudo-spin couplings of both Zeeman and spin-orbit character.
Ingenious combinations of nano-magnetic and nano-electronic
fabrication capabilities can thus be used to realize new device
architectures that are impossible to achieve in other materials.
Additional variability is provided by the valley-isospin degree of
freedom, allowing the selective creation of global or
valley-contrasting real-spin or pseudo-spin polarizations and their
exploitation for both fundamental research and new nano-spintronic
applications.
Finally, it is fascinating that the phenomena discussed here are
present not only in bilayer graphene, but they exist in any material
with similar symmetries. In particular, analogous behaviour can be
realized in appropriately designed metamaterials~\cite{gib09, gom12}
or cold atoms in optical lattices~\cite{tar12}. The
intrinsic tuneability of system parameters in these latter
realizations opens intriguing avenues to tailor the relative
magnitude of these matter-field interactions beyond the regime
accessible in BLG.

\acknowledgments The authors thank M.~Meyer and H.~Schulthei{\ss}
for help with generating Fig.~\ref{fig:visual}. Useful discussions
with J.~J.~Heremans, A.~H.~MacDonald, J.~L.\ Ma\~nes, and
M.~Morgenstern are also gratefully acknowledged. This work was
supported by Marsden Fund contract no.\ VUW0719, administered by the
Royal Society of New Zealand, and at Argonne National Laboratory by
the DOE BES via contract no.\ DE-AC02-06-CH11357.


\appendix

\section{Symmetry analysis and invariant expansion for BLG bandstructure:
General discussion and results} \label{app:A}

The Hamiltonian (\ref{eq:effM_BLG}) describes the band structure of
graphene as a function of the wave vector $\kk$ measured from the
$\vek{K}$ point. The theory of invariants \cite{bir74, win10a}
makes it possible to also include perturbations $\vekc{K}$
that are combinations of various quantities in addition to the wave
vector $\kk$, e.g., electric and magnetic fields $\vekc{E}$ and
$\vek{B}$, strain $\Strain$ and the intrinsic spin $\vek{s}$. The
electron states at the $\vek{K}$ point transform according to the
two-dimensional irreducible representation (IR) $\Gamma_3$ of the
group $D_3$ (Ref.~\onlinecite{kos63}). The effective $2 \times 2$
Hamiltonian for the $\Gamma_3$ subspace can be expressed as
\begin{equation}
  \label{eq:invar}
  \mathcal{H}_\vek{K} (\vekc{K})
  =  \sum_{\kappa, \, \lambda}
  \koeff{a}{}{\kappa\lambda}
  \sum_{l=1}^{L_\kappa} {X}_l^{(\kappa)}
  \mathcal{K}_l^{(\kappa,\lambda) \, \ast} \, .
\end{equation}
Here $\koeff{a}{}{\kappa\lambda}$ are prefactors, ${X}_l^{(\kappa)}$
are $2\times 2$ matrices that transform according to the IRs
$\Gamma_\kappa$ (of dimension $L_\kappa$) contained in the product
representation $\Gamma_3 \times \Gamma_3^\ast = \Gamma_1 + \Gamma_2
+ \Gamma_3$ of $D_3$. Likewise, $\vekc{K}$ can be decomposed into
irreducible tensor operators $\vekc{K}^{(\kappa',\lambda)}$ that
transform according to the IRs $\Gamma_{\kappa'}$ of $D_3$. Using
the coordinate system in Fig.~\ref{fig:lattice} we obtain the basis
matrices and tensor operators listed in Tables~\ref{tab:basemat} and
\ref{tab:tensorop}.


\begin{table}[b]
  \caption[]{\label{tab:basemat} Symmetrized matrices for the invariant
   expansion of the block $\mathcal{H}_{33}$ for the point group
   $D_{3}$.}
$\arraycolsep 0.8em
\renewcommand{\arraystretch}{1.2}
\begin{array}{clcl} \hline \hline
\mbox{Block} &
\multicolumn{1}{l}{\mbox{Representations}} &
\multicolumn{2}{c}{\mbox{Symmetrized matrices}} \\ \hline
\mathcal{H}_{11} & \Gamma_1 \times \Gamma_1^\ast = \Gamma_1 &
    \Gamma_1: & (1)  \\
\mathcal{H}_{22} & \Gamma_2 \times \Gamma_2^\ast = \Gamma_1 &
    \Gamma_1: & (1)  \\
\mathcal{H}_{13} & \Gamma_1 \times \Gamma_3^\ast = \Gamma_3 &
    \Gamma_3: & (1,1), (-i,i)  \\
\mathcal{H}_{23} & \Gamma_2 \times \Gamma_3^\ast = \Gamma_3 &
    \Gamma_3: & (1,-1), (-i,-i)  \\
\mathcal{H}_{33} & \Gamma_3 \times \Gamma_3^\ast &
    \Gamma_1: & \openone  \\
& = \Gamma_1 + \Gamma_2 + \Gamma_3
  & \Gamma_2: & \sigma_z \\
  & & \Gamma_3: & \sigma_x, \sigma_y
\\ \hline \hline
\end{array}$
\end{table}

\begin{table}[b]
  \caption[]{\label{tab:tensorop}
   Irreducible tensor components for the point group $D_{3}$ (the
   group of the $\vek{K}$ point in BLG). Terms printed in bold give
   rise to invariants in $\mathcal{H}_\vek{K} (\vekc{K})$
   allowed by time-reversal invariance. (No terms proportional to
   $k_z$ are listed as they are irrelevant for graphene.)
   Contributions that are new in BLG (i.e., terms not allowed for
   the group $D_{3h}$ of the $\vek{K}$ point in SLG~\cite{win10a})
   are shown in \textcolor{red}{red}.
   Notation: $\{A, B\} \equiv \frac{1}{2} (AB + BA)$.}
  \begin{textmath}
  \extrarowheight 0.2ex
  \newcommand{\nl}{\newline}
  \begin{array}{cs{1.0em}p{0.9\columnwidth}} \hline \hline
    \Gamma_1 & \allowed{1};
    \allowed{k_x^2 + k_y^2};
    \allowed{\{k_x, 3k_y^2 - k_x^2\}};
    \forbidden{B_x k_x + B_y k_y}; \nl
    \forbidden{k_x \Ee_x + k_y \Ee_y};
    \allowednew{\Ee_x B_x + \Ee_y B_y};
    \allowednew{\Ee_z B_z};
    \allowed{\strain_{xx} + \strain_{yy}};
    \allowednew{\strain_{zz}}; \nl
    \allowed{(\strain_{yy} - \strain_{xx}) k_x + 2 \strain_{xy} k_y};
    \allowednew{\strain_{yz} k_x - \strain_{xz} k_y}; \nl
    \forbidden{(\strain_{yy} - \strain_{xx}) B_x + 2 \strain_{xy} B_y};
    \forbidden{\strain_{yz} B_x - \strain_{xz} B_y}; \nl
    \forbidden{(\strain_{yy} - \strain_{xx}) \Ee_x + 2 \strain_{xy} \Ee_y};
    \forbidden{\strain_{yz} \Ee_x - \strain_{xz} \Ee_y};
    \forbidden{s_x k_x + s_y k_y}; \nl
    \allowed{s_x B_x + s_y B_y};
    \allowed{s_z B_z};
    \allowednew{s_x \Ee_x + s_y \Ee_y};
    \allowednew{s_z \Ee_z}; \nl
    \allowed{(s_x k_y - s_y k_x) \Ee_z};
    \allowed{s_z (k_x \Ee_y - k_y \Ee_x)}; \nl
    \forbidden{s_x (\strain_{yy} - \strain_{xx}) + 2 s_y \strain_{xy}};
    \forbidden{s_x \strain_{yz} - s_y \strain_{xz}} \\
    \Gamma_2 & \forbidden{\{k_y, 3k_x^2-k_y^2\}};
    \allowed{B_z};
    \allowednew{k_x B_y - k_y B_x};
    \allowednew{\Ee_z};
    \allowed{k_x \Ee_y - k_y \Ee_x}; \nl
    \forbidden{\Ee_x B_y - \Ee_y B_x};
    \forbidden{(\strain_{xx} - \strain_{yy}) k_y + 2 \strain_{xy} k_x};
    \forbidden{\strain_{yz} k_y + \strain_{xz} k_x}; \nl
    \allowednew{(\strain_{xx} - \strain_{yy}) B_y + 2 \strain_{xy} B_x};
    \allowed{(\strain_{xx} + \strain_{yy}) B_z};
    \allowednew{\strain_{zz} B_z}; \nl
    \allowednew{\strain_{xz} B_x + \strain_{yz} B_y};
    \allowed{(\strain_{xx} - \strain_{yy}) \Ee_y + 2 \strain_{xy} \Ee_x}; \nl
    \allowednew{(\strain_{xx} + \strain_{yy}) \Ee_z};
    \allowednew{\strain_{zz} \Ee_z};
    \allowednew{\strain_{xz} \Ee_x + \strain_{yz} \Ee_y};
    \allowed{s_z}; \nl
    \allowednew{s_x k_y - s_y k_x};
    \forbidden{s_x B_y - s_y B_x};
    \forbidden{s_x \Ee_y - s_y \Ee_x}; \nl
    \forbidden{(s_x k_x + s_y k_y) \Ee_z};
    \allowednew{s_y (\strain_{xx} - \strain_{yy}) + 2 s_x \strain_{xy}};\nl
    \allowed{s_z (\strain_{xx} + \strain_{yy})};
    \allowednew{s_x \strain_{xz} + s_y \strain_{yz}};
    \allowednew{s_z \strain_{zz}}; \\
    %
    \Gamma_3 &
    \allowed{k_x, k_y};
    \allowed{\{k_y+k_x, k_y - k_x\}, 2 \{k_x, k_y\}}; \nl
    \allowed{\{k_x, k_x^2 + k_y^2\}, \{k_y, k_x^2 + k_y^2\}}; \nl
    \forbidden{B_x, B_y};
    \forbidden{B_y k_y - B_x k_x, B_x k_y + B_y k_x};
    \forbidden{B_z k_y, - B_z k_x}; \nl
    \forbidden{\Ee_x, \Ee_y};
    \forbidden{\Ee_y k_y - \Ee_x k_x, \Ee_x k_y +  \Ee_y k_x};
    \forbidden{\Ee_z k_y, - \Ee_z k_x}; \nl
    \allowednew{\Ee_y B_y - \Ee_x B_x, \Ee_y B_x + \Ee_x B_y};
    \allowed{\Ee_y B_z, - \Ee_x B_z}; \nl
    \allowed{\Ee_z B_y, - \Ee_z B_x};
    \allowed{\strain_{yy} - \strain_{xx}, 2 \strain_{xy}};
    \allowednew{\strain_{yz}, - \strain_{xz}}; \nl
    \allowed{(\strain_{xx} + \strain_{yy}) (k_x, k_y)};
    \allowednew{\strain_{yz} k_x + \strain_{xz} k_y,
     \strain_{xz} k_x  - \strain_{yz} k_y}; \nl
    \allowed{(\strain_{xx} - \strain_{yy}) k_x + 2\strain_{xy} k_y,
     (\strain_{yy} - \strain_{xx}) k_y + 2\strain_{xy} k_x}; \nl
    \allowednew{\strain_{zz} k_x, \strain_{zz} k_y};
    \nl
    \forbidden{(\strain_{xx} + \strain_{yy}) (B_x, B_y)};
    \forbidden{\strain_{yz} B_x + \strain_{xz} B_y,
     \strain_{xz} B_x  - \strain_{yz} B_y};  \nl
    \forbidden{(\strain_{xx} - \strain_{yy}) B_x + 2 \strain_{xy} B_y,
     (\strain_{yy} - \strain_{xx}) B_y + 2 \strain_{xy} B_x}; \nl
    \forbidden{2 \strain_{xy} B_z, (\strain_{xx} - \strain_{yy}) B_z};
    \forbidden{\strain_{zz} B_x, \strain_{zz} B_y};
    \forbidden{\strain_{xz} B_z, \strain_{yz} B_z}; \nl
    \forbidden{(\strain_{xx} + \strain_{yy}) (\Ee_x, \Ee_y)};
    \forbidden{\strain_{yz} \Ee_x + \strain_{xz} \Ee_y,
     \strain_{xz} \Ee_x  - \strain_{yz} \Ee_y}; \nl
    \forbidden{(\strain_{xx} - \strain_{yy}) \Ee_x + 2 \strain_{xy} \Ee_y,
     (\strain_{yy} - \strain_{xx}) \Ee_y + 2 \strain_{xy} \Ee_x}; \nl
    \forbidden{2 \strain_{xy} \Ee_z, (\strain_{xx} - \strain_{yy}) \Ee_z};
    \forbidden{\strain_{zz} \Ee_x, \strain_{zz} \Ee_y};
    \forbidden{\strain_{xz} \Ee_z, \strain_{yz} \Ee_z}; \nl
    \forbidden{s_x, s_y};
    \forbidden{s_y k_y - s_x k_x, s_x k_y + s_y k_x};
    \forbidden{s_z k_y, - s_z k_x}; \nl
    \allowed{s_y B_y - s_x B_x, s_x B_y + s_y B_x};
    \allowednew{s_z B_y, - s_z B_x}; \nl
    \allowednew{s_y B_z, - s_x B_z};
    \allowednew{s_y \Ee_y - s_x \Ee_x, s_x \Ee_y + s_y \Ee_x}; \nl
    \allowed{s_z \Ee_y, - s_z \Ee_x};
    \allowed{s_y \Ee_z, - s_x \Ee_z}; \nl
    \allowed{s_z (k_x \Ee_y + k_y \Ee_x), s_z (k_x \Ee_x - k_y \Ee_y)}; \nl
    \allowed{(s_x k_y + s_y k_x) \Ee_z, (s_x k_x - s_y k_y) \Ee_z}; \nl
    \forbidden{(s_x, s_y) (\strain_{xx} + \strain_{yy})};
    \forbidden{2 s_z \strain_{xy}, s_z (\strain_{xx} - \strain_{yy})}; \nl
    \forbidden{s_x (\strain_{xx} - \strain_{yy}) - 2 s_y \strain_{xy},
     s_y (\strain_{yy} - \strain_{xx}) - 2 s_x \strain_{xy}}; \nl
    \forbidden{s_x \strain_{zz}, s_y \strain_{zz}};
    \forbidden{s_z \strain_{xz}, s_z \strain_{yz}};
    \forbidden{s_x \strain_{yz} + s_y \strain_{xz},
     s_x \strain_{xz}  - s_y \strain_{yz}};
    \\ \hline \hline
  \end{array}
  \end{textmath}
\end{table}


Additional constraints for the Hamiltonian (\ref{eq:invar}) are due
to time reversal invariance. The point group $D_{3d}$ of BLG
contains symmetry elements $R$ mapping the basis functions
$\Psi_{\vek{K} \lambda}$ at $\vek{K}$ on $\Psi_{\vek{K}' \lambda}$
at $\vek{K}'$. These basis functions are also mapped onto each other
by the time-reversal operation $\theta$, i.e., we have
\begin{equation}
  \label{eq:time}
  \theta \, \Psi_{\vek{K},\lambda} = \Psi_{\vek{K}\lambda}^\ast
  = \sum_{\lambda'} \mathcal{T}_{\lambda\lambda'}
  \: \Psi_{\vek{K}'\lambda'} \,,
\end{equation}
with a unitary matrix~$\mathcal{T}$. Combining these operations, we
obtain \cite{bir74, man07, win10a}
\begin{equation}
  \label{eq:timea2}
  \mathcal{T}^{-1} \mathcal{H}_\vek{K} (R^{-1} \vekc{K}) \mathcal{T}
  = \mathcal{H}_\vek{K}^\ast (\zeta \vekc{K})
  = \mathcal{H}_\vek{K}^t (\zeta \vekc{K}) .
\end{equation}
Here $t$ denotes transposition and $\zeta$ depends on the behavior
of $\vekc{K}$ under time reversal. $\kk$, $\vek{B}$, and $\vek{s}$
are odd under time reversal so that then $\zeta=-1$, while
$\vekc{E}$ and $\Strain$ have $\zeta=+1$. Equation (\ref{eq:timea2})
provides a general criterion for determining which terms in the
expansion (\ref{eq:invar}) are allowed by time-reversal invariance
and which terms are forbidden. The matrix~$\mathcal{T}$ depends on
the choice for the operation $R$. If $R$ is the reflection $R_y$ at
the $yz$ plane [thus mapping the atoms in each sublattice in each
layer onto each other, see Fig.~\ref{fig:lattice}(a)], we obtain
\begin{equation}\label{eq:reflect}
  \mathcal{H}_{\vek{K}'} (\vekc{K}) = \mathcal{H}_\vek{K} (R_y^{-1}
  \vekc{K}) \quad .
\end{equation}
and the matrix~$\mathcal{T}$ is simply the identity matrix.

We note that under $R_y$ polar ($\vek{p}$) and axial ($\vek{a}$)
vectors transform as
\begin{subequations}
  \label{eq:transVal}
  \begin{align}
    \label{eq:transPol}
    p_x & \to -p_x \quad \, & \quad p_{y, z} & \to p_{y, z} \, \quad ,\\
    \label{eq:transAxi}
    a_x & \to a_x \quad \, & \quad a_{y, z} & \to -a _{y, z} \, \quad .
  \end{align}
\end{subequations}
The transformational properties for the components of the
second-rank strain tensor $\strain_{ij}$ can be expressed
similarly~\cite{bir74}. If we denote by $\tau_0$ ($\tau_z$) the
unity (diagonal Pauli) matrix acting in valley-isospin space,
Eq.~(\ref{eq:transVal}) can be summarized by writing general vector
operators as $(p_x\,\tau_z, p_y\, \tau_0, p_z\,\tau_0)$ and $(a_x\,
\tau_0, a_y\, \tau_z, a_z\, \tau_z)$.

The group $D_3$ characterizing the $\vek{K}$ point in BLG is a
subgroup of the group $D_{3h}$ for the $\vek{K}$ point in SLG
\cite{kos63} so that any term allowed by spatial symmetries in
$\mathcal{H}_\vek{K} (\vekc{K})$ for SLG is likewise allowed in BLG.
Moreover, the constraint (\ref{eq:timea2}) due to time reversal
invariance is exactly equivalent to the constraint in SLG
\cite{win10a}. Thus it follows immediately that the invariant
expansion for BLG contains all terms that exist already for SLG.

A more detailed analysis shows that the point group $D_{3h}$ for SLG
distinguishes, as is usual, between polar vectors (such as the
electric field $\vekc{E}$) and axial vectors (such as the magnetic
field $\vek{B}$). Thus each term in the SLG Hamiltonian with a
certain functional form and linear in the field $\vekc{E}$ or
$\vek{B}$ is forbidden for the other field. However, the point group
$D_3 \subset D_{3h}$ relevant for BLG contains only rotations as
symmetry elements so that it cannot distinguish between polar and
axial vectors. Therefore, the $x$ and $y$ components of any vector
transform according to the IR $\Gamma_3$, whereas the $z$ component
transforms according to $\Gamma_2$. This implies that spatial
symmetries cannot distinguish electric and magnetic fields in BLG.
Moreover, Eq.\ (\ref{eq:timea2}) treats electric and magnetic fields
symmetrically, too. Thus it follows that every $\vekc{E}$-dependent
term in the BLG Hamiltonian (\ref{eq:invar}) is accompanied by
another term where $\vekc{E}$ is simply replaced by $\vek{B}$ (and
vice versa for $\vek{B}$-dependent terms). However, the prefactors
of these terms are, in general, unrelated~\cite{bir74}.
Table~\ref{tab:newTerms} summarizes some of the new terms arising
from this magneto-electric equivalence.

Similar to SLG, pseudospin up and down in BLG corresponds to atoms
in sublattice $B$ and $B'$, see
Fig.~\ref{fig:lattice}(a)~\cite{mcc57}. With the convention that for
both valleys $v = \vek{K}, \vek{K}'$ the up and down eigenstates of
pseudospin $\sigma_z$ correspond to the same sublattice, the above
procedure determines the Hamiltonians $\mathcal{H}_v$ up to phases
$\phi_v$ corresponding to pseudospin rotations $\tilde{\sigma}_j =
\exp(i\phi_v \sigma_z/2) \, \sigma_j \exp(-i\phi_v \sigma_z/2)$ of
the basis matrices $\sigma_j$ ($j=x,y$) entering the invariant
expansion (\ref{eq:invar}) (Ref.~\onlinecite{voo09}). Due to this
$U(1)$ gauge freedom the direction of the average in-plane spin
polarization $\braket{\vek{\sigma}_\|}_v$ induced within a valley by
fields $\vek{B}$ and $\vekc{E}$ is well-defined only up to a uniform
field-independent angular offset $\phi_v$.

\section{Valley-contrasting axion electrodynamics in BLG:
Calculation of induced densities and currents}\label{app:B}

The presence of $\mathcal{L}_\mathrm{ax}$ leads to modifications of
the inhomogeneous Maxwell's equations (i.e., Gauss' and Amp\`ere's
laws) amounting to the replacements~\cite{wil87}
\begin{subequations}
\label{eq:macroscopic}
\begin{eqnarray}
\vek{D} &\to& \vek{D} - \sum_\alpha \nu_\alpha \left(c_\| \,
\vek{B}_\|+ c_z\, B_z \, \hat{\vek{z}}\right) \quad , \\
\vek{H} &\to& \vek{H} + \sum_\alpha \nu_\alpha \left(c_\| \,
\vek{\Ee}_\| + c_z\, \Ee_z \, \hat{\vek{z}}\right) \quad ,
\end{eqnarray}
\end{subequations}
where $\vek{D}$ and $\vek{H}$ denote the macroscopic electric and
magnetic fields in matter~\cite{jac99}. Equations
(\ref{eq:macroscopic}) imply the existence of extra valley-dependent
charge densities and currents that are most generally given by
\begin{subequations}
\begin{eqnarray}
\rho^{(\alpha)}  &=&  \nu_\alpha \vek{\nabla}\cdot\left(c_\|\,
\vek{B}_\| + c_z \, B_z\,\hat{\vek{z}}\right)  , \\
\vek{j}^{(\alpha)} &=& - \nu_\alpha \vek{\nabla} \times\left(c_\|\,
\vek{\Ee}_\| + c_z \, \Ee_z\hat{\vek{z}}\right) \nonumber \\ &&
- \nu_\alpha\, \partial_t \left(c_\|\, \vek{B}_\| +
c_z \, B_z\,\hat{\vek{z}}\right) . \,\,\,\,
\end{eqnarray}
\end{subequations}

The coefficients $c_j$ are finite constants within BLG but
must vanish outside the sample. Hence, there is a boundary
contribution to $\vek{\nabla} c_j \equiv
\vek{\nabla}_\| c_j$, and we have
\begin{equation}
\rho^{(\alpha)}  =  \nu_\alpha \left[ \left(\vek{\nabla}_\| c_\|
\right)\cdot\vek{B}_\| + c_\|\,\vek{\nabla}\cdot\vek{B}_\|
+ c_z \,\partial_z B_z \right]  .
\end{equation}
Using the homogeneous Maxwell equation $\vek{\nabla}\cdot\vek{B}=0$,
the magnetic-field-induced charge densities are then given by
Eq.~(\ref{eq:axion_rho}). Furthermore, the two-dimensionality of
BLG implies that currents flow only in-plane. Hence, both
$(\vek{\nabla}_\| c_\|)\times\vek{\Ee}_\|$ and $\partial_t B_z
\hat{\vek{z}}$ do not contribute to the current, which is just given
by
\begin{equation}\label{eq:j_intermed}
\vek{j}^{(\alpha)} = - \nu_\alpha\left[ c_\|\left( \vek{\nabla}\times
\vek{\Ee}_\| +\partial_t \vek{B}_\| \right) + \vek{\nabla}\times \left(
c_z\, \Ee_z\,\hat{\vek{z}}\right)\right] \, .
\end{equation}
The other homogeneous Maxwell equation can be written as
$\vek{\nabla} \times \vek{\Ee}_\| + \partial_t \vek{B}_\| =
-\vek{\nabla} \times \left(\Ee_z\, \hat{\vek{z}}\right) -
\partial_t \, B_z\,\hat{\vek{z}}$ whose in-plane projection inserted
into Eq.~(\ref{eq:j_intermed}) yields Eq.~(\ref{eq:axion_j}). For
homogeneous, time-independent fields $\vekc{E}$ and $\vek{B}$ the
ordinary couplings of $\vekc{E}$ and $\vek{B}$ to $\rho$ and
$\vek{j}$ vanish so that we only get the valley-contrasting axion
electrodynamics of Eqs.\ (\ref{eq:axion_rho}) and
(\ref{eq:axion_j}).


The valley-contrasting axion electrodynamics discussed here can be
expected to apply in situations when the valley-isospin degree of
freedom is conserved. This will generally be the case for
inhomogeneities induced by smoothly varying external fields (such as
the one proposed in Ref.~\onlinecite{mar08}). In contrast, the
physical termination of the BLG sheet is abrupt on an atomic scale,
and details will usually matter~\cite{li10}. Nevertheless, the
recently observed~\cite{li11a} emergence of a universal behavior at
disordered BLG edges consistent with that obtained from our
valley-contrasting axion-electrodynamics suggests a wider
applicability of this formalism.

\section{Magnitude of Prefactors in Eq.~(\ref{eq:fieldHam})}
\label{app:C}

Recent experiments~\cite{zha09} demonstrated that a displacement
field of $\sim 1$~V/nm generates a bandgap of $\sim 0.1$~eV in
BLG, implying that $\lambda_z \sim 0.1$~$e\,$nm (with electron
charge $e$) consistent with first-principles
calculations~\cite{kon12}. For the remaining prefactors in
Eq.~(\ref{eq:fieldHam}), the Slonczewski-Weiss-McClure (SWM)
model~\cite{mcc57} applied to BLG yields (see also
Ref.~\onlinecite{zha11})
\begin{eqnarray}
  g &=& \frac{4\gamma_{0}\gamma_4}{\gamma_1 \gamma_a}
  \approx 6.2 \; , \\
  c_z &=& \frac{\gamma_0^2 + \gamma_4^2}{\gamma_1^2} \,
  \frac{\mu_{\text{B}}\, \lambda_z}{2 \, \gamma_a}
  \approx 3 \times 10^{-4}~e\, \mathrm{nm / T} \; .
\end{eqnarray}
The SWM parameters $\gamma_0$, $\gamma_1$, and $\gamma_4$ are
defined in Ref.~\onlinecite{mcc57}, and we have $\gamma_a =
2\hbar^2/ (3 m_0 a^2)$, where $m_0$ is the electron mass in vacuum
and $a$ is the lattice constant of BLG's planar honeycomb structure.


\end{document}